\documentclass[twocolumn,preprintnumbers,amsmath,amssymb,superscriptaddress]{revtex4-2}
\usepackage[utf8]{inputenc}
\usepackage{latexsym,amssymb,amsthm,amsmath,epsfig}
\usepackage[left=2.00cm, right=2.00cm, top=2.00cm, bottom=2.00cm]{geometry}
\usepackage{float}
\usepackage{tabularx}
\usepackage{comment}
\usepackage{hyperref}

\hypersetup{
    citecolor=cyan,
    colorlinks=true,
    linkcolor=blue,
    filecolor=blue,      
    urlcolor=blue,
}

\begin{document}
\title{Feedback-enhanced distant entanglement of magnon and phonon modes with atomic ensembles in coupled cavities}

\author{Muhammad Awais Altaf}
\altaffiliation{Corresponding author: \href{altafawais9@gmail.com}{altafawais9@gmail.com}}
\affiliation{Department of Physics, University of Mianwali, Mianwali $42200$, Pakistan}
\author{Muhammad Irfan}
\affiliation{Department of Physics and Applied Mathematics, Pakistan Institute of Engineering and Applied Science (PIEAS), Nilore, Islamabad $45650$, Pakistan}
\affiliation{Center for Mathematical Sciences, PIEAS, Nilore, Islamabad $45650$, Pakistan}

\date{\today }

\begin{abstract}
The generation and manipulation of distant entanglement between disparate systems is crucial for various quantum technologies.
In this work, we investigate a system of coupled cavities comprising an ensemble in cavity-1, a yttrium-iron-garnet (YIG) sphere in cavity-2, and a coherent feedback loop (CFL) that feeds the output of cavity-1 back into cavity-1 through a beam splitter.
This system features five excitation modes: cavity-1 photons, atomic ensemble, cavity-2 photons, the magnon, and phonon modes of the YIG sphere.
Thus various combinations of bipartite entanglements can be studied.
Our main focus is the study of various combinations of distant bipartite entanglements, especially the entanglement of the atomic ensemble and photons in cavity-1 with the magnon and phonon modes of the YIG sphere in cavity-2.
Compared to the previously reported results, introducing a CFL significantly enhances all the bipartite entanglements.
Besides, the degree of entanglement of various modes, the parameters space, where the strong entanglement exists, is also significantly enhanced due to CFL.
Moreover, the entanglement is more robust against thermal noise.
We believe our results are important for quantum technologies where the distribution of entanglement on quantum networks is crucial.
\end{abstract}

\maketitle

\section{introduction}
Quantum entanglement, a fundamental concept in quantum mechanics, refers to the strong correlations between quantum systems such that one system's state instantaneously influences another's state, regardless of the distance between them~\cite{PhysRev.47.777, RevModPhys.81.865}.
This non-local connection is the most important resource that enables various quantum technologies including quantum computing, quantum sensing, and quantum communication.
For instance, entanglement enhances quantum key distribution in quantum communication and provides additional layers of security~\cite{PhysRevLett.84.4729, PhysRevA.63.012309, PhysRevA.78.032314, 10.1126/sciadv.abe6379}.
Quantum internet is another potential technology that exploits the quantum entanglement to outperform the traditional communication systems~\cite{kimble2008quantum, doi:10.1126/science.aam9288}.
However, deploying such technologies on a global scale requires reliable quantum networks, which crucially depend on the realization of quantum repeaters~\cite{RevModPhys.95.045006}.
Among many potential candidates for the realization of quantum repeaters, the atomic ensemble-based repeaters received significant interest in recent years due to their longer coherence times, making them reliable memory nodes~\cite{RevModPhys.83.33}.

The integration of atomic ensembles with various macroscopic quantum systems has been a topic of fundamental interest.
This includes quantum entanglement between the collective excitations of atomic ensembles and the mechanical modes of an oscillator~\cite{PhysRevA.77.050307, bai2016robust, PhysRevA.96.042320}.
For instance, Genes \textit{et. al.} demonstrated the existence of the atom-field-mirror entanglement in a cavity optomechanical system~\cite{PhysRevA.77.050307}.
It is also shown that atomic ensembles can be entangled with phonon modes placed in distant cavities~\cite{bai2016robust}.

Quantum entanglement between atomic ensembles and magnonics systems has recently received considerable attention.
In this regard, cavity magnomechanics is a rapidly advancing field, where a magnomechanical system comprising a magnon and a mechanical mode (phonon) couples to a microwave cavity mode~\cite{doi:10.1126/sciadv.1501286}, offering a tripartite interaction within a single device.
Experimental studies have demonstrated the possibility of strong coupling between the magnons and cavity photons\cite{zhang2015cavity, PhysRevLett.113.156401}.
Li \textit{et. al.} showed the existence of bipartite and tripartite entanglement in such a cavity magnomechanical system having a yttrium iron garnet (YIG) sphere inside a microwave cavity~\cite{PhysRevLett.121.203601}, followed by proposals to enhance the entanglement using squeezing and coherent feedback loop (CFL)~\cite{PhysRevA.105.063704, Ding:22, Sohail:23, amazioug2023enhancement, https://doi.org/10.1002/qute.202400281, Lin_CFL-JPA, Amazioug2025}.
The initial proposal was later extended to more complex hybrid systems~\cite{PhysRevA.104.023711, Li_2021, wang2022nonreciprocal}.
Recently, magnons have been integrated with cavity optomechanics, allowing the entanglement of magnons, phonons, and photons in the optical regime~\cite{PhysRevB.108.024105, Di:23, https://doi.org/10.1002/lpor.202200866}.

The entanglement between the atomic ensembles and magnon modes has recently been studied in cavity magnomechanical and optomagnomechanical systems~\cite{PhysRevApplied.20.034043, PhysRevA.108.023501, PhysRevA.109.043708, Di:24, MA2024130899, Di:24, Hidki2024}.
For instance, Wu \textit{et. al.} examined the entanglement of an atomic ensemble and a magnon mode in an optical cavity and found that stronger magnetic dipole coupling or the magnetostrictive coupling leads to stronger entanglement~\cite{PhysRevApplied.20.034043}.
Li \textit{et. al.} explored entanglement in an optomagnomechanical system \cite{PhysRevA.108.023501}.
Moreover, Dilawaiz \textit{et. al.} studied distant ensemble-magnon entanglement in a coupled microwave cavity system~\cite{PhysRevA.109.043708}.
Ma \textit{et. al.} considered a microwave cavity coupled with an optomechanical cavity to study ensemble-magnon entanglement~\cite{MA2024130899}. Sohail \textit{et. al.} studied the distant magnon-magnon entanglement via photon hopping in coupled microwave cavities~\cite{Sohail2023}.

In this paper, we study a system of two coupled microwave cavities, one with an atomic ensemble of two-level atoms (typically $N\sim10^7$ \cite{PhysRevA.77.050307, PhysRevLett.83.1319}) and the other a $250~\mu$m YIG sphere \cite{doi:10.1126/sciadv.1501286}.
A microwave field drives cavity-1, whereas another field drives the YIG sphere in cavity-2.
We employ a CFL such that the output of cavity-1 is injected back into cavity-1 via a beam splitter which also injects the microwave field in cavity-1.
Previous studies have shown that CFL, under the appropriate conditions, improves entanglement in cavity optomechanics and magnomechanics~\cite{PhysRevA.95.043819, amazioug2023enhancement, LAKHFIF2024129678}.
We show that the presence of CFL enhances all types of distant as well as other entanglements among various degrees of freedom like atomic ensemble, cavity-1 and 2 photons, phonon and magnon modes of YIG sphere~\cite{PhysRevA.109.043708}.
More importantly, the parameter space for entanglement is significantly expanded in the presence of the CFL.
We also show a significant increase in the robustness of entanglement against temperature due to the CFL.
We focus on studying distant bipartite entanglement between macroscopic degrees of freedom, including ensemble-magnon and ensemble-phonon entanglement.
Studying such distant entanglement is crucial not only for testing fundamental theories but also for having potential applications in quantum communication and hybrid quantum networks.
We believe the results presented in this work may find important applications in quantum communication networks where atomic ensemble-based entanglement plays a crucial role.

\section{System Model and Hamiltonian}

\begin{figure*}
\centering
\includegraphics[width=7.0 in]{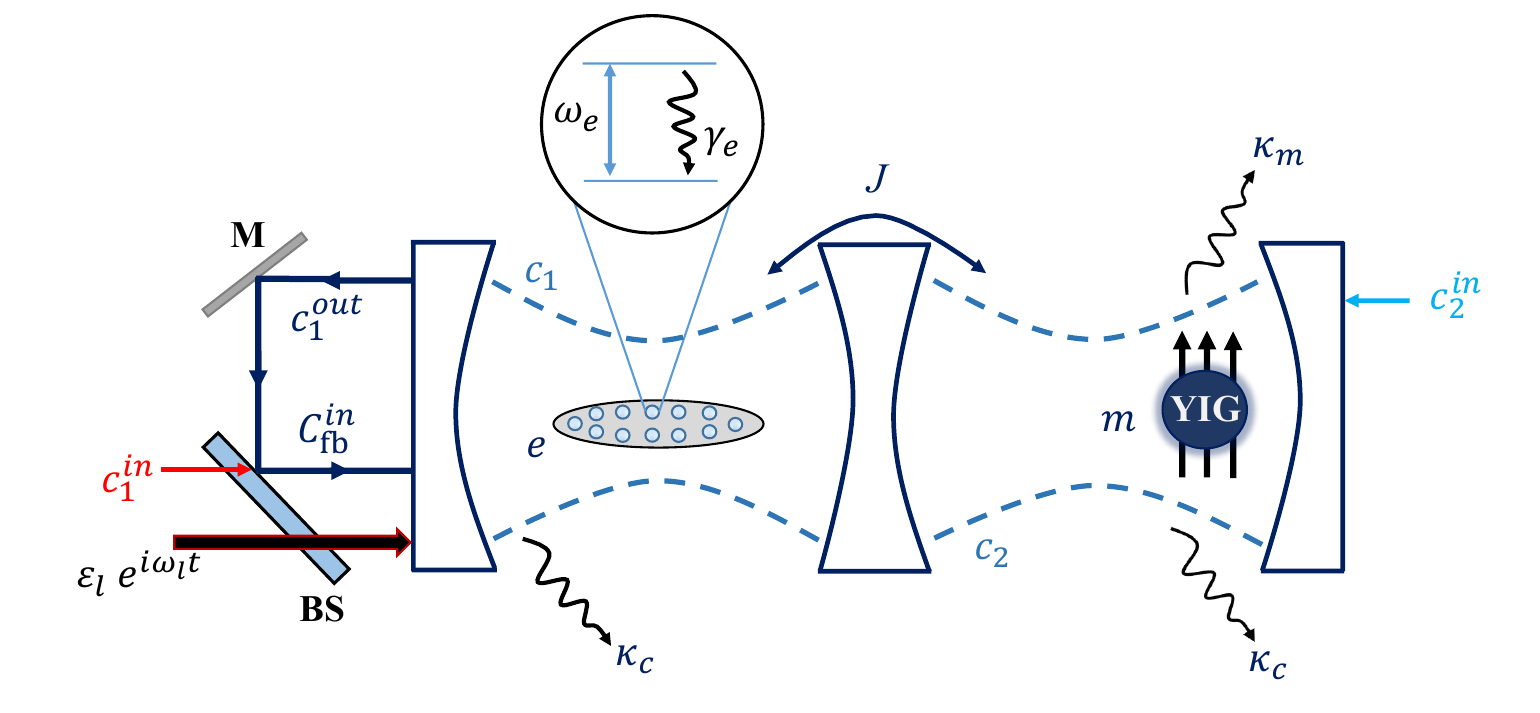}
\caption{Schematic diagram of the coupled microwave cavities in the presence of CFL. cavity-1 contains an ensemble of $N$ two-level atoms with excitation frequency $\omega_e$, and a YIG sphere is present in cavity-2. Both cavities are coupled to each other with coupling strength $J$. A microwave source with frequency $\omega_l$ and amplitude $\varepsilon_l$ is injected into cavity-1 using a beam splitter (BS) of adjustable reflectivity $r_B$ and transmissivity $t_B$. While the magnon modes of the YIG sphere are driven by $\omega_l$ with strength $\varepsilon_m$ (not shown). The beam splitter also serves another important purpose: it feeds the output of cavity-1 ($c_1^{out}$) back into cavity-1 with the help of a highly reflecting mirror (M). The decay rates of cavities modes ($c_1$ and $c_2$), magnon ($m$), ensemble ($e$), and phonon ($b$) modes are $\kappa_c, \kappa_m$, $\gamma_e$, and $\gamma_b$ respectively. The noise entering cavity-1 through the beam splitter is $c_1^{in}$ (but it is modified to $c_{\text{fb}}^{in}$ due to the CFL (see text for details)), CFL-modified input field is represented by $C_{\text{fb}}^{in}$, and noise $c_2^{in}$ enters through the right mirror of cavity-2.}
\label{fig:sch}
\end{figure*}

We consider a system that consists of two coupled microwave cavities with coupling strength $J$ as shown in Fig.~\ref{fig:sch}.
cavity-1 contains an atomic ensemble of $N$ two-level atoms having transition frequency $\omega_e$, while a YIG sphere is placed in cavity-2.
An external bias magnetic field \textbf{H} is applied on the YIG sphere in the $z$ direction that excites the magnon modes.
The magnon mode frequency $\omega_m$ is related to the gyromagnetic ratio ($\gamma$) and bias magnetic field as $\omega_m=\gamma$\textbf{H}.
The change in magnetic field leads to mechanical deformation due to magnetostriction in the YIG sphere, resulting in phonon modes.
Additionally, we employ a beam splitter in front of cavity-1 serving two purposes: (i) it transmits the cavity-1 driving field with frequency $\omega_l$, and (ii) it re-injects the cavity-1 output field into cavity-1 to implement CFL. 

To describe the ensemble of $N$ two-level atoms, individually characterized by spin-1/2 Pauli algebra ($\sigma_z, \sigma_{\pm}$, with $\sigma_+(\sigma_-)$, the raising (lowering) spin-operators), we adopt the collective spin operators $S_{\pm,z}=\sum_{j=1}^{N} \sigma^{(j)}_{\pm, z}$, following the commutation relations  $[S_+, S_-]=S_z$ and $[S_z, S_\pm]=\pm2 S_\pm$ \cite{bai2016robust}.
If we assume the population of atoms in the ground state is much larger than the excited state population, we obtain $S_z \simeq \langle S_z \rangle \simeq -N$ \cite{PhysRevA.77.050307}.
Under this condition, we can apply the Holstein-Primakoff transformation \cite{PhysRev.58.1098} to write the collective spin operators in terms of bosonic creation (annihilation) operator $e^\dagger$($e$) as: $S_+=e^\dagger \sqrt{N-e^\dagger e} \simeq \sqrt{N}e^\dagger$, $S_-= \sqrt{N-e^\dagger ee} \simeq \sqrt{N}e$, and $S_z=e^\dagger e -N/2$ such that $[e, e^\dagger]=1$. 

The Hamiltonian of the system under rotating-wave approximation in a frame rotating with the frequency of the drive field is given by ($\hbar=1$)
\begin{equation}
    \mathcal{H} = \mathcal{H}_{0} + \mathcal{H}_{I} + \mathcal{H}_{D},
    \label{eq1}
\end{equation}
    
where
\begin{eqnarray}
    \mathcal{H}_0 &=& \sum_{l=1}^{2} \Delta_l c_l^\dagger c_l + \Delta_m m^\dagger m + \Delta_e e^\dagger e  + \frac{\omega_b}{2} (q^2 + p^2),\nonumber \\
    \mathcal{H}_I &=& g_{mb} m^\dagger m q +g_{mc}(c_2m^{\dag}+ c_2^{\dag}m) + J (c_1^\dagger c_2 + c_1 c_2^\dagger)\nonumber \\ 
    &+& G_{ce} (e c_1^\dagger + e^\dagger c_1),\nonumber \\
    \mathcal{H}_D &=& i t_B \varepsilon_l (c_1^\dagger - c_1) + i \varepsilon_m (m^\dagger - m).
    \label{eq2}
\end{eqnarray}
Here, the first term $\mathcal{H}_0$ in Eq. \ref{eq2} is the free Hamiltonian of cavity-1 and cavity-2, atomic ensemble, and magnon modes with annihilation (creation) operators $c_l(c_l^{\dag})$, $e(e^{\dag})$, and $m(m^{\dag})$, respectively.
It contains the phonon mode with frequency $\omega_b$, where $q$ and $p$ are the dimensionless position and momentum quadratures of the phonon mode, respectively.
The second term $\mathcal{H}_I$ is the interaction Hamiltonian that includes the magnon-phonon, cavity-2 photon-magnon, cavity-cavity, and cavity-1 photon-ensemble couplings which are denoted by $g_{mb}$, $g_{mc}$, $J$, and $G_{ce}$, respectively.
The last term $\mathcal{H}_D$ is the Hamiltonian of the driving fields.
It consists of a microwave field that drives cavity-1 with amplitude $\varepsilon_l = \sqrt{2P\kappa_c/\hbar \omega_l}$ with input power $P$ of the drive field and $\kappa_c$ the decay rate of cavity-1.
The parameter $t_B$ represents the transmissivity of the beam splitter such that $r_B^2+t_B^2=1$ \cite{Mandel1995OpticalCA}, with $r_B$ its reflectivity.
A second field having Rabi frequency $\varepsilon_m= \frac{5}{4}\gamma\sqrt{N_s}B_0$ drives the magnon mode, where $B_0$ is the amplitude of the applied magnetic field and $N_s$ represents the total number of spins.
Here, in the case of YIG sphere, $\gamma = 2\pi \times 28$ GHz/T, $N_s = \rho V$ with spin density $\rho =4.22 \times 10^{27}$ m$^{-3}$ and volume $V$ of the sphere \cite{PhysRevLett.121.203601}.

As previously mentioned, the feedback channel re-injects the output field of cavity-1 via a beam splitter. Using the standard input-output relation~\cite{amazioug2023enhancement, PhysRevA.30.1386}, the cavity-1 output field is given by: $c_1^{out}=\sqrt{2\kappa_c} c_1 - t_B c_1^{in}$.
The beam splitter mixes the input noise and feedback field such that the modified input field is given by $C^{in}_{\text{fb}}=t_B c_1^{in} +  r_B e^{i\phi} c_1^{out}$ with the first term representing a part of the input noise of cavity-1 while the second term indicates a portion of the output field of cavity-1 ($c_1^{out}$) that the beam splitter reflects and allows to enter the cavity-1 \cite{PhysRevA.95.043819}.
Therefore, the bare input noise $c_1^{in}$, in the absence of CFL is modified by the $C^{in}_{\text{fb}}$ with the resulting  CFL-modified input noise operator of cavity-1: $c^{in}_{\text{fb}}=t_B(1- r_B e^{i\phi}) c_1^{in}$.
To study the dynamics of the system in the presence of CFL, we write the following Quantum Langevin Equations (QLEs):
\begin{eqnarray}
    \dot{c}_1 &=& -(\kappa_{\text{fb}} + i\Delta_{\text{fb}}) c_1 - i G_{ce} e - i J c_2 + t_B \varepsilon_l + \sqrt{2\kappa_c} c_{\text{fb}}^{in}, \nonumber \\
    \dot{c}_2 &=& -(\kappa_c + i\Delta_2) c_2 - i J c_1 - i g_{mc} m + \sqrt{2\kappa_c} c_2^{in}, \nonumber \\
    \dot{e} &=& -(\gamma_e + i\Delta_e) e - i G_{ce} c_1 + \sqrt{2\gamma_e} e^{in}, \nonumber \\
    \dot{m} &=& - (\kappa_m + i \Delta_m) m - i g_{mc} c_2 - i g_{mb} m q + \varepsilon_m\nonumber \\
    &+& \sqrt{2\kappa_m} m^{in}, \quad \dot{q} = \omega_b p,\nonumber\\
    \dot{p} &=& -\omega_b q - \gamma_b p - g_{mb} m^\dagger m + \zeta.
    \label{eq3}
\end{eqnarray}
The presence of $c_1$ term in $C^{in}_{\text{fb}}$ (due to CFL) modifies the cavity decay rate and detuning~\cite{PhysRevA.95.043819}.
The effective decay rate is given by $\kappa_{\text{fb}}=\kappa_c(1-2r_B \cos\phi)$, where $\kappa_c$ is the original decay rate of the cavity, $r_B$ is the reflection coefficient of the beam splitter, and $\phi$ is the phase shift of the output field.
Similarly, the detuning is modified as $\Delta_{\text{fb}}=\Delta_1$ $- 2\kappa_c r_B \sin\phi$, where $\Delta_1$ is the original detuning.
These expressions indicate that the feedback can either suppress or enhance the cavity decay rate and detuning depending on the phase shift.
The decay rates of cavity-2 and magnon modes are represented by $\kappa_c$ and $\kappa_m$, respectively, whereas the decay rates of the atomic ensemble and phonon modes are denoted by $\gamma_e$ and $\gamma_b$, respectively. 

Since the effective input noise is also modified due to the CFL, it leads to modified correlation functions of the noise operator $c^{in}_{\text{fb}}$ in the time domain~\cite{amazioug2023enhancement, e25101462}: $\langle c_{\text{fb}}^{in}(t) c_{\text{fb}}^{in \dag}(t')\rangle = t_B^2|1 - r_B e^{i\phi}|^2 [W_{c_1}(\omega_{c_1}) + 1]\delta(t-t')$, $\langle c_{\text{fb}}^{in \dag}(t) c_{\text{fb}}^{in}(t')\rangle= t_B^2|1 - r_B e^{i\phi}|^2 W_{c_1}(\omega_{c_1})\delta(t-t')$.
Therefore, by tuning the CFL parameters, we can control the correlations of the input noise operators.

On the other hand, $c_2^{in}$, $e^{in}$, $m^{in}$, and $\zeta$ are the input noise operators for the cavity-2, ensemble, magnon and phonon modes, respectively. Their noise operators are characterized by the correlation functions in the time domain \cite{RevModPhys.82.1155} as
$\langle c_2^{in}(t) c_2^{in \dag}(t')\rangle$ $= [W_{c_2}(\omega_{c_2}) + 1 ]\delta(t-t')$, $\langle c_2^{in \dag}(t) c_2^{in}(t')\rangle = W_{c_2}(\omega_{c_2})\delta(t-t')$, $\langle e^{in}(t) e^{in \dag}(t')\rangle = \delta(t-t')$, $\langle m^{in}(t)$ $m^{in \dag}(t')\rangle= \left[W_m(\omega_m) + 1\right]\delta(t-t')$, $\langle m^{in \dag}(t) m^{in}(t')\rangle = W_m(\omega_m)\delta(t-t')$, and $\langle \zeta(t) \zeta(t') + \zeta(t')\zeta(t)\rangle/2 = \gamma_d\left[2W_b(\omega_b) + 1\right]\delta(t-t')$.

Here $W_j(\omega_j) = [\text{exp}(\hbar \omega_j/k_B T)-1]^{-1} (j = c_1, c_2,m,b)$ are the equilibrium mean number of the thermal photon, magnon, and phonon at bath temperature $T$ with $k_B$ the Boltzmann constant.

Under the strong cavity and magnon drive assumption, we obtain large steady-state amplitude $|c_{ls}| \gg1 (l=1,2)$ and $|m_s| \gg1$.
As previously noted, the bosonic representation of an atomic ensemble is valid under a low excitation limit.
The condition  $g^2/(\Delta_e^2 + \gamma_e^2) \ll |c_{1s}|^{-2} \ll 1$ simultaneously satisfy both low-excitation and large steady-state amplitude limits.
Therefore, we can linearize the dynamics of the system around the steady-state values by writing the operators as $o_s$ and small 1st-order fluctuations $\delta o$ (neglecting second-order fluctuation terms) with zero mean value: $o = o_s + \delta o$ $(o \in c_l, e, m, q, p)$.
Hence, the steady-state values of cavities, ensemble, magnon, and phonon mode operators are written as:
 
\begin{eqnarray}
c_{1s} &=& \frac{t_B\varepsilon_l -iG_{ce} e_s -iJ c_{2s}}{\kappa_{\text{fb}} + i\Delta_{\text{fb}}},\hspace{0.1cm} c_{2s} = \frac{-i(J c_{1s} + g_{mc} m_s)}{\kappa_c + i\Delta_2}, \nonumber\\
e_s &=& \frac{-i G_{ce} c_{1s}} {\gamma_e + i\Delta_e}, \quad m_s  = \frac{\varepsilon_m - ig_{mc}  c_{2s}} {\kappa_m+i\tilde{\Delta}_m},\nonumber \\
q_s &=& -(g_{mb}/\omega_b)| m_s|^2, \quad p_s = 0,
\label{eq4}
\end{eqnarray}
where $\tilde{\Delta}_m=\Delta_m + g_{mb} q_s$ is the effective magnon detuning including the frequency shift due to the magnomechanical interaction while the effective magnomechanical coupling rate is $G_{mb} = i \sqrt{2}g_{mb} m_\text{s}$.
We define the quadrature of fluctuation operators $\delta A_o = (\delta o + \delta o^{\dag})/\sqrt{2}$ and $\delta B_o = i(\delta o^{\dag} - \delta o)/\sqrt{2}$ \cite{PhysRevA.82.012333}.
We write the linearized quantum Langevin equations for quadrature fluctuation operators as: 
\begin{equation}
    \Dot{\mathcal{F}}(t) = \mathcal{A F}(t) + n(t),
    \label{eq5}
\end{equation}
where $\mathcal{F}(t)$ = [$\delta A_{c_1}(t)$, $\delta B_{c_1}(t)$, $\delta A_{c_2}(t)$, $\delta B_{c_2}(t)$, $\delta A_{m}(t)$, $\delta B_{m}(t)$, $\delta q(t)$, $\delta p(t)$, $\delta A_{e}(t)$, $\delta B_{e}(t)]^T$ is the fluctuation operator in the form of quadrature fluctuation operators and the noise column vector is $n(t) =$ [$\sqrt{2\kappa_c} A_{c_1}^{in}(t)$, $\sqrt{2\kappa_c} B_{c_1}^{in}(t)$, $\sqrt{2\kappa_c} A_{c_2}^{in}(t)$, $\sqrt{2\kappa_c} B_{c_2}^{in}(t)$, $\sqrt{2\kappa_m}$ $A_m^{in}(t), \sqrt{2\kappa_m} B_m^{in}(t)$, 0, $\zeta$, $\sqrt{2\gamma_e}$ $A_e^{in}(t)$, $\sqrt{2\gamma_e} B_e^{in}(t)]^T$. The drift matrix $\mathcal{A}$ is given by:

\begin{widetext}
    \begin{equation}
    \mathcal{A} = 
    \begin{pmatrix}
    -\kappa_{\text{fb}} & \Delta_{\text{fb}} & 0 & J & 0 & 0 & 0 & 0 & 0 & G_{ce}\\
    -\Delta_{\text{fb}} & -\kappa_{\text{fb}} & -J & 0 & 0 & 0 & 0 & 0 & -G_{ce} & 0\\
    0 & J & -\kappa_c & \Delta_2 & 0 & g_{mc} & 0 & 0 & 0 & 0\\
    -J & 0 & -\Delta_2 & -\kappa_c & -g_{mc} & 0 & 0 & 0 & 0 & 0\\
    0 & 0 & 0 & g_{mc} & -\kappa_m & \tilde{\Delta}_m & -G_{mb} & 0 & 0 & 0\\
    0 & 0 & -g_{mc} & 0 & -\tilde{\Delta}_m & -\kappa_m & 0 & 0 & 0 & 0\\
    0 & 0 & 0 & 0 & 0 & 0 & 0 & \omega_b & 0 & 0\\
    0 & 0 & 0 & 0 & 0 & G_{mb} & -\omega_b & -\gamma_b & 0 & 0\\
    0 & G_{ce} & 0 & 0 & 0 & 0 & 0 & 0 & -\gamma_e & \Delta_e\\
    -G_{ce} & 0 & 0 & 0 & 0 & 0 & 0 & 0 & -\Delta_e & -\gamma_e\\
    \end{pmatrix}.  
    \label{eq6}
    \end{equation}
\end{widetext}
The linearized quantum Langevin equations are fundamental in analyzing the stability of quantum systems. These equations can be associated with an effective linearized Hamiltonian, which plays a crucial role in maintaining the Gaussian state of the system under stable conditions. Therefore, due to the linearized dynamics and the Gaussian nature of the quantum noises, the system naturally evolves into a continuous-variable (CV) five-mode Gaussian state. To assess the stability of the linearized system, the Routh-Hurwitz criterion is employed \cite{PhysRevA.35.5288}. This criterion provides a mathematical framework to determine the stability conditions by analyzing the eigenvalues of the drift matrix ($\mathcal{A}$). The system is considered stable and can thus achieve a steady state only when real parts of all eigenvalues of $\mathcal{A}$ are negative. Moreover, the variance within each subsystem and the covariance across several subsystems are described by the Covariance Matrix ($\mathcal{C}$), which is generated from the Lyapunov equation when the stability requirements are fulfilled \cite{doi:10.1080/03043799308928354};
\begin{equation}
    \mathcal{A C} + \mathcal{C} \mathcal{A}^T + \mathcal{D}=0,
    \label{eq7}
\end{equation}
where $\mathcal{D}$=diag[$\kappa_c t_B^2|1 - r_B e^{i\phi}|^2(2W_{c1}+1)$, $\kappa_c t_B^2|1 - r_B e^{i\phi}|^2(2W_{c1}+1)$, $\kappa_c (2W_{c2}+1)$, $\kappa_c (2W_{c2}+1)$, $\kappa_m(2W_{m}+1)$, $\kappa_m(2W_{m}+1)$, 0, $\gamma_b(2W_{b}+1)$, $\gamma_e$, $\gamma_e]^T$ is the diffusion matrix. 
This five-mode Gaussian state system is characterized by a covariance matrix $\mathcal{C}$,
\begin{equation}
    \mathcal{C} = 
    \begin{pmatrix}
    \mathcal{C}_{c_1} & \mathcal{C}_{c_1c_2} & \mathcal{C}_{c_1m} & \mathcal{C}_{c_1b} & \mathcal{C}_{c_1e}\\
    \mathcal{C}^T_{c_1c_2} &\mathcal{C}_{c_2} & \mathcal{C}_{c_2m} & \mathcal{C}_{c_2b} & \mathcal{C}_{c_2e}\\
    \mathcal{C}^T_{c_1m} & \mathcal{C}^T_{c_2m} & \mathcal{C}_{m} & \mathcal{C}_{mb} & \mathcal{C}_{me}\\
    \mathcal{C}^T_{c_1b} & \mathcal{C}^T_{c_2b} & \mathcal{C}^T_{mb} & \mathcal{C}_{b} & \mathcal{C}_{be}\\
    \mathcal{C}^T_{c_1e} & \mathcal{C}^T_{c_2e} & \mathcal{C}^T_{me} & \mathcal{C}^T_{be} & \mathcal{C}_{e}
\end{pmatrix}.   
\label{eq8}
\end{equation} 
Each block is a $2\times2$ matrix in which diagonal blocks represent the variance inside each subsystem, i.e., cavity-1 photon, cavity-2 photon, magnon, phonon, and ensemble. On the other hand, the off-diagonal blocks specify the covariances across the subsystems \cite{bai2016robust}.
To study the bipartite entanglement of two subsystems, the $10\times10$ covariance matrix $\mathcal{C}$ must be reduced to $\mathcal{C}_l$, a $4\times4$ matrix.
For example, the first four rows and columns of $\mathcal{C}$ determine the sub-matrix that describes the covariance of the cavity-1 photon and cavity-2 photon subsystems \cite{bai2016robust},

\begin{equation}
    C_l = 
    \begin{pmatrix}
   \mathcal{C}_{c_1} & \mathcal{C}_{c_1c_2}\\
    \mathcal{C}^T_{c_1c_2} &\mathcal{C}_{c_2}
\end{pmatrix}.
\label{eq9}
\end{equation}
This also applies to the other subsystems. To analyze the bipartite entanglement among different subsystems of the coupled two-cavity system, we use logarithmic negativity ($E_N$) \cite{PhysRevA.65.032314, PhysRevLett.95.090503}:
\begin{equation}
    E_N = \text{max}[0, \text{ln} (2 \xi^-)],
    \label{eq10}
\end{equation}
where $\xi^-$ = min eig $|i\oplus_{n=1}^2 (i\sigma_y) \mathcal{\tilde{C}}|$ is the minimum symplectic eigenvalue of the reduced covariance matrix $\mathcal{\tilde{C}}=\mathcal{T}_{1,2} \mathcal{C}_l \mathcal{T}_{1,2}$. Here $\sigma_y$ is the Pauli y-matrix and partial transposition matrix is $\mathcal{T}_{1,2}=\text{diag}(1,-1,1,1)$ \cite{PhysRevLett.84.2726}.

\section{results and discussion}
\begin{figure}
\centering
\includegraphics[width=3.40 in]{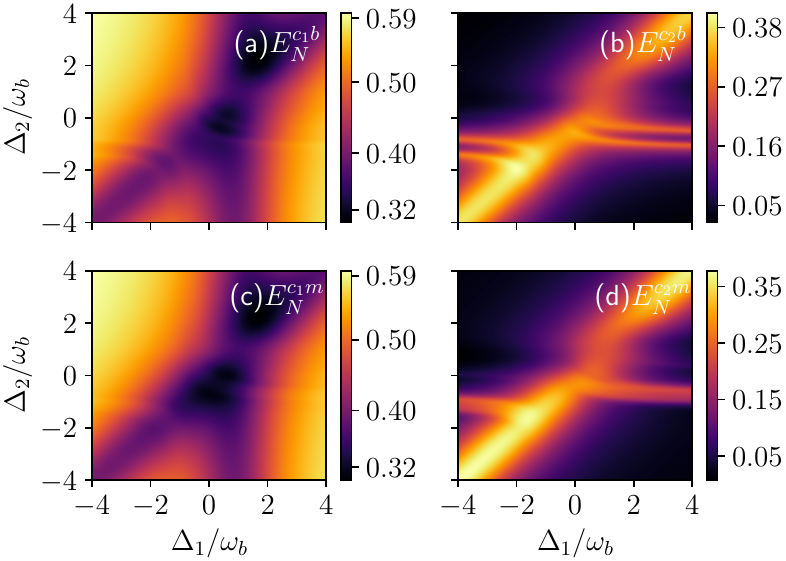}
\caption{Density plots of entanglement between (a) cavity-1 photon and phonon $E_N^{c_1b}$, (b) cavity-2 photon and phonon $E_N^{c_2b}$, (c) cavity-1 photon and magnon $E_N^{c_1m}$ and (d) cavity-2 photon and magnon $E_N^{c_2m}$ subsystems as a function of cavity-1 detuning $\Delta_1/\omega_b$ and cavity-2 detuning $\Delta_2/\omega_b$. Here we choose the value of $\Delta_e/\omega_b=1.0$, $\Tilde{\Delta}_m/\omega_b=0.9$, and $J/\omega_b=0.8$.}
\label{f1}
\end{figure}

Now we discuss the numerical results of our simulations.
We use the following experimentally suitable parameters~\cite{PhysRevLett.121.203601}: $\kappa_c/2\pi=\kappa_m/2\pi=1$MHz, $\omega_l/2\pi=\omega_m/2\pi=10$GHz, $\omega_b/2\pi=10$MHz, $\gamma_b/2\pi=100$Hz, $\gamma_e/2\pi=1$MHz, $g_{mc}/2\pi=3.2$MHz, $G_{mb}/2\pi=4.8$MHz, $G_{ce}/2\pi=6$MHz, and $T=10$mK.
We chose the beam splitter reflectivity $r_B=0.75$ and phase $\phi=\pi$ to obtain strong entanglement.
In Fig. \ref{f1}, we show the bipartite entanglement of (a) cavity-1 photon-phonon ($E^{c_1b}_N$), (b) cavity-2 photon-phonon ($E^{c_2b}_N$), (c) cavity-1 photon-magnon ($E^{c_1m}_N$), and (d) cavity-2 photon-magnon ($E^{c_2m}_N$) subsystems as a function of dimensionless detunings $\Delta_1/\omega_b$ and $\Delta_2/\omega_b$.
We choose ensemble detuning $\Delta_e/\omega_b=1.0$, effective magnon detuning $\Tilde{\Delta}_m/\omega_b=0.9$, and cavities coupling strength $J/\omega_b=0.8$.
The result shows that we have significant $E^{c_1b}_N$ and $E^{c_1m}_N$ entanglements for the complete range of considered detunings with maximum values around $(\Delta_1, \Delta_2) \approx (\mp 4.0\omega_b, \pm 4.0\omega_b)$.
A comparison with the earlier study~\cite{PhysRevA.109.043708} shows that these distant entanglements are significantly enhanced in the presence of CFL.
The cavity-2 photon-phonon $E^{c_2b}_N$ and cavity-2 photon-magnon $E^{c_2m}_N$ subsystems are strongly entangled for $\Delta_1 \approx \Delta_2$ with a maximum value around $\Delta_1 \approx \Delta_2 \approx -2\omega_b$.

\begin{figure}
\centering
\includegraphics[width=3.40 in]{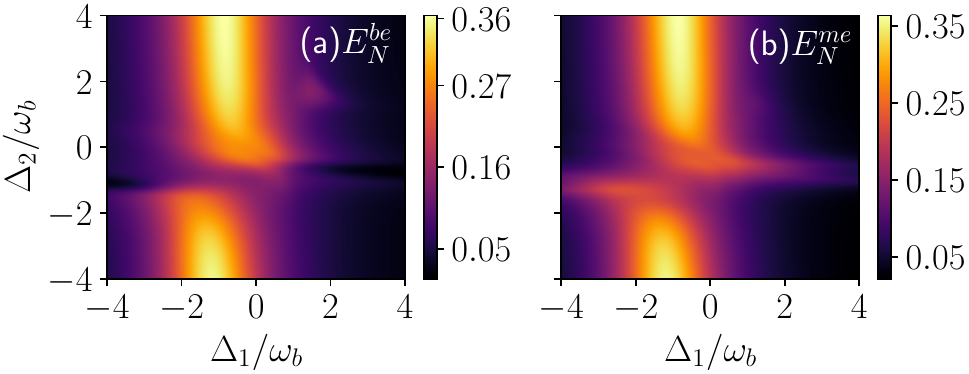}
\caption{Density plots of entanglement of (a) phonon-ensemble $E_N^{be}$ and (b) magnon-ensemble $E_N^{me}$ subsystems as a function of cavity-1 detuning $\Delta_1/\omega_b$ and cavity-2 detuning $\Delta_2/\omega_b$. We select $\Delta_e/\omega_b=-1.0$ and other parameters are same as in Fig. \ref{f1}.}
\label{f2}
\end{figure}

Next, we plot the macroscopic distant entanglement of the atomic ensemble (in cavity-1) with phonon and magnon modes (in cavity-2).
In Fig. \ref{f2} (a) $E^{be}_N$ and (b) $E^{me}_N$ are presented as a function of $\Delta_1/\omega_b$ and $\Delta_2/\omega_b$.
We use $\Delta_e/\omega_b=-1.0$, whereas the rest of the parameters are the same as in Fig.\ref{f1}.
It can be seen that $E^{be}_N$ is strongly entangled around $\Delta_1 \approx -\omega_b$ for almost all the values of $\Delta_2$ except region around $ -\omega_b$, reaching maximum values at $\Delta_2 \approx \pm 4\omega_b$.
Similarly, $E^{me}_N$ reaches its peak entanglement at nearly the same values of $\Delta_1$ and $\Delta_2$ as $E^{be}_N$.
Importantly, the CFL-induced enhancement is significant as compared to the earlier study~\cite{PhysRevA.109.043708}.

\begin{figure}[t]
\centering
\includegraphics[width=0.50 \textwidth]{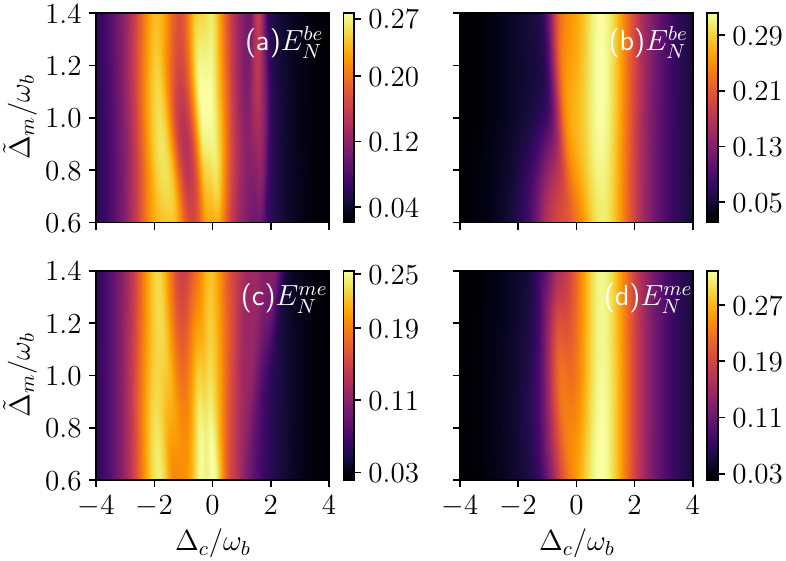}
\caption{Density plots (a)-(b) of $E_N^{be}$ and (c)-(d) of $E_N^{me}$ as a function of normalized cavity detuning $\Delta_c/\omega_b$ and magnon detuning $\tilde{\Delta}_m/\omega_b$. Here (a) and (c) are plotted for the symmetric detuning $\Delta_c/\omega_b=\Delta_1/\omega_b=\Delta_2/\omega_b$. On the other hand, (b) and (d) are for non-symmetric detuning $\Delta_c/\omega_b=-\Delta_1/\omega_b=\Delta_2/\omega_b$. Remaining parameters are $\Delta_e/\omega_b=-1$, and $J/\omega_b=1.0$.}
\label{f5}
\end{figure}
Since the choice and sign of both cavity detunings play crucial roles, we consider two distinct scenarios.
The first scenario is the symmetric detuning where both cavities have the same detuning ($\Delta_1=\Delta_2=\Delta_c$), which means if the first cavity is red-detuned or blue-detuned, then so is the second cavity.
The second scenario is the non-symmetric case, where cavity-1 and cavity-2 have opposite detunings ($\Delta_1=-\Delta_2=-\Delta_c$).
It demonstrates that if the first cavity is red-detuned, then the second is blue-detuned, and vice versa.
In Fig. \ref{f5}, we show the bipartite entanglement between phonon-ensemble ($E^{be}_N$) and magnon-ensemble ($E^{me}_N$) versus normalized cavity detuning $\Delta_c/\omega_b$ and effective magnon detuning $\tilde{\Delta}_m/\omega_b$ assuming $\Delta_e/\omega_b=-1.0$ and $J/\omega_b=1.0$.
In Fig. \ref{f5} (a) and (c) symmetric detuning case is presented, where $E^{be}_N$ and $E^{me}_N$ both approach maximum around resonance ($\Delta_c \approx0$).
Both types of entanglement are significant for the whole range of effective magnon detuning with optimum values of $\tilde{\Delta}_m/\omega_b\approx1.2$ and $\tilde{\Delta}_m/\omega_b\approx0.6$ for $E^{be}_N$ and $E^{me}_N$, respectively.
For the non-symmetric detuning case shown in Fig. \ref{f5} (b) and (d), both types of entanglements are significant around $\Delta_c \approx \omega_b$ for the whole range of $\tilde{\Delta}_m$.

\begin{figure}
\centering
\includegraphics[width=0.50 \textwidth]{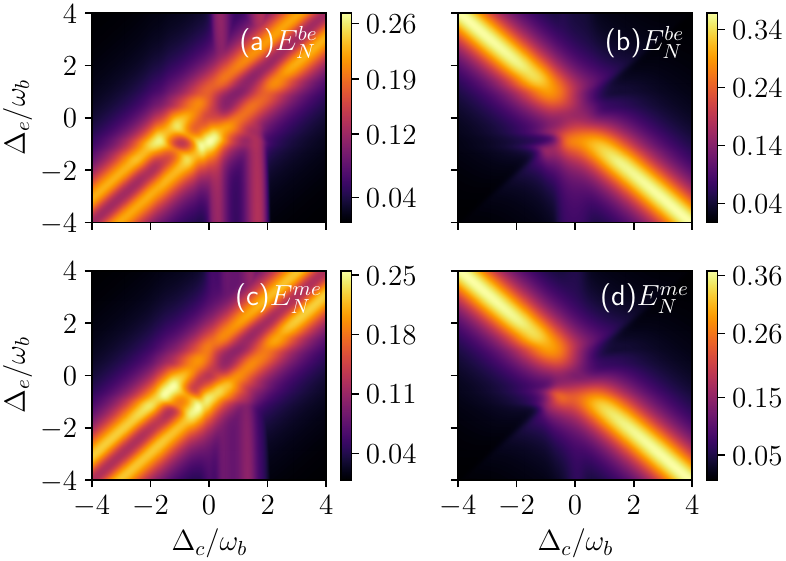}
\caption{Density plots (a)-(b) of $E_N^{be}$ and (c)-(d) of $E_N^{me}$ as a function of normalized cavity detuning $\Delta_c/\omega_b$ and ensemble detuning $\Delta_e/\omega_b$. Here (a) and (c) are plotted when $\Delta_c/\omega_b=\Delta_1/\omega_b=\Delta_2/\omega_b$, while (b) and (d) are for $\Delta_c/\omega_b=-\Delta_1/\omega_b=\Delta_2/\omega_b$. Other parameters are $\tilde{\Delta}_m/\omega_b=0.9$, and $J/\omega_b=0.8$.}
\label{f6}
\end{figure}

In Fig. \ref{f6}, we illustrate the entanglement of phonon-ensemble and magnon-ensemble subsystems as a function of $\Delta_c/\omega_b$ and $\Delta_e/\omega_b$ for the symmetric case (left panel) and the non-symmetric case (right panel), assuming $\tilde{\Delta}_m/\omega_b=0.9$ and $J/\omega_b=0.8$.
A comparison of both panels shows a very distinctive behavior between the two cases.
For the symmetric case, we obtain significant phonon-ensemble entanglement at $\Delta_c/\omega_b \approx0$ and $\Delta_e/\omega_b \approx-2$ whereas it attains 
maximum entanglement at two points, one around $\Delta_c/\omega_b \approx-1$ and $\Delta_e/\omega_b \approx-3$ and second in the vicinity of $\Delta_c/\omega_b \approx2$ and $\Delta_e/\omega_b \approx-4$ for the non-symmetric case.
The magnon-ensemble entanglement is significant when $\Delta_c \approx \Delta_e$ for the symmetric case with maximum values around $\Delta_c/\omega_b \approx-1.5$ and $\Delta_e/\omega_b \approx-1.5$.
For the non-symmetric case, $E_N^{me}$ is significant when $\Delta_c\approx-\Delta_e$.
In all situations, the entanglement is significantly improved as compared to the case when no CFL is considered.

\begin{figure}
\centering
\includegraphics[width=3.35 in]{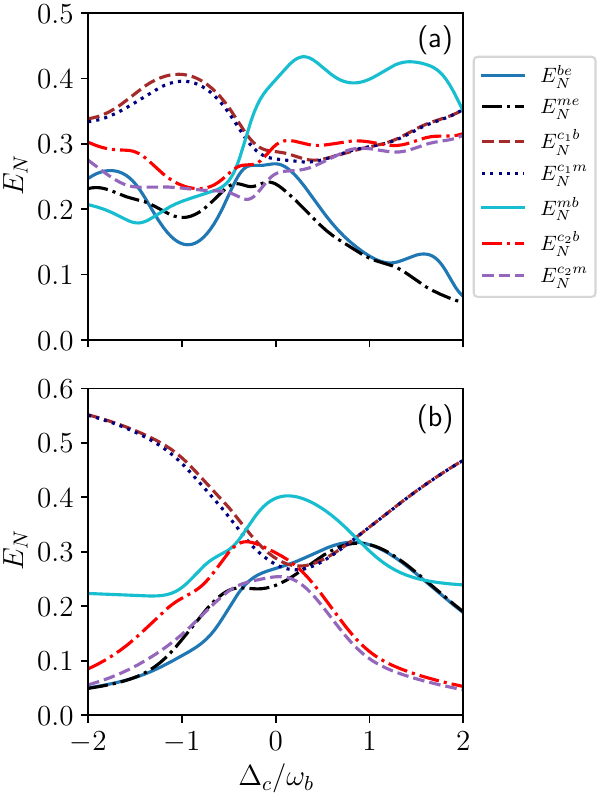}
\caption{The bipartite entanglements $E_{N}^{be}$, $E_{N}^{me}$, $E_{N}^{c_1b}$, $E_{N}^{c_1m}$, $E_{N}^{mb}$, $E_{N}^{c_2b}$, and $E_{N}^{c_2m}$ as a function of normalized cavity detuning $\Delta_c/\omega_b$ at $J/\omega_b=0.4$ when (a) $\Delta_c/\omega_b=\Delta_1/\omega_b=\Delta_2/\omega_b$ and (b) $\Delta_c/\omega_b=-\Delta_1/\omega_b=\Delta_2/\omega_b$. Here $\Delta_e/\omega_b=-1.0$, and $\tilde{\Delta}_m/\omega_b=0.9$.}
\label{f7}
\end{figure}


Next in Fig. \ref{f7}, we present the line plots of the entanglement of seven subsystems, $E_N^{be}$, $E_N^{me}$, $E_N^{c_1b}$, $E_N^{c_1m}$, $E_N^{mb}$, $E_N^{c_2b}$ and $E_N^{c_2m}$, as a function of normalized cavity detuning $\Delta_c/\omega_b$ at $\Delta_e/\omega_b=-1$, $\tilde{\Delta}_m/\omega_b=0.9$, and $J/\omega_b=0.4$.
The left panel is for the symmetric detuning while the right is for the non-symmetric case.
We note that cavity-1 photon-phonon ($E_N^{c_1b}$) and cavity-1 photon-magnon ($E_N^{c_1m}$) remain strongly entangled for the whole range of detuning for both symmetric and non-symmetric cases.
The distant macroscopic entanglement ($E_N^{be}$ and $E_N^{me}$) shows variable entanglement for the whole range of considered frequencies with a maximum value around $\Delta_c\approx-\omega_b$ for the symmetric case and $\Delta_c \approx \omega_b$ for the non-symmetric case.
The entanglement of magnon and phonon modes with the cavity-2 photons ($E_N^{c_2b}$ and $E_N^{c_2m}$) shows variation as a function of detuning.
For the symmetric case, it remains very significant, while for the non-symmetric case, the entanglement dies out for $-\omega_b \lessapprox \Delta_c\gtrapprox\omega_b$.
Finally, the phonon-magnon entanglement shows a pronounced peak at  $\Delta_c \approx 0.8\omega_b$ for the symmetric case, while maintaining a high degree of entanglement across the frequency spectrum for both symmetric and non-symmetric detunings.
The variations in entanglement show the transfer of entanglement across various subsystems as a function of the detuning.
Importantly, all types of entanglement show significant CFL-induced enhancement in the degree of entanglement compared to the earlier study~\cite{PhysRevA.109.043708}.

\begin{figure}
\centering
\includegraphics[width=3.40 in]{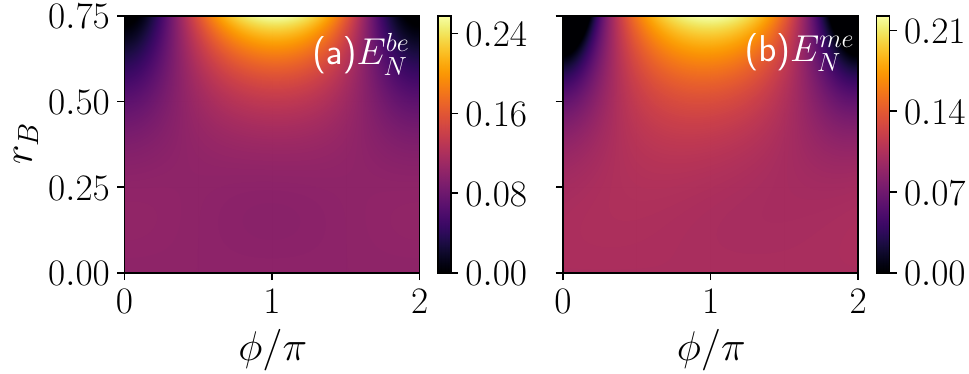}
\caption{Density plots of entanglement of (a) phonon-ensemble $E_N^{be}$ and (b) magnon-ensemble $E_N^{me}$ subsystems as a function of reflectivity $r_B$ and phase shift $\phi/\pi$. In (a) $\Delta_1/\omega_b=\Delta_2/\omega_b=-1.75$ and (b) $\Delta_1/\omega_b=-2.0$ and $\Delta_2/\omega_b=-1.40$. Other parameters are $\Delta_e/\omega_b=-1.0$, $\Tilde{\Delta}_m/\omega_b=0.9$, and $J/\omega_b=0.8$. }
\label{f9}
\end{figure}

\begin{table}[h]
    \centering
    \begin{tabular}{|c|c|c|c|c|c|c|}
        \hline
        \textbf{Entanglement} &$\boldsymbol{\Delta_1/\omega_b}$ & $\boldsymbol{\Delta_2/\omega_b}$ & $\boldsymbol{\Delta_e/\omega_b}$ & $\boldsymbol{\tilde{\Delta}_m/\omega_b}$ & $\boldsymbol{J/\omega_b}$ \\ \hline
        $E_{N}^{be}$ & $-0.88  $ & $4.0  $ & $-1.0  $ & $0.9  $ & $0.8  $ \\ \hline
        $E_{N}^{me}$ & $ -0.88  $ & $4.0 $ & $-1.0  $ & $0.9  $ & $0.8  $ \\ \hline
        $E_{N}^{c_1b}$ & $-4.0  $ & $4.0  $ & $1.0  $ & $0.9  $ & $0.8  $ \\ \hline
        $E_{N}^{c_1m}$ & $-4.0  $ & $4.0  $ & $1.0  $ & $0.9  $ & $0.8  $ \\ \hline
    \end{tabular}
    \caption{Set of parameters that are used in Fig. \ref{f8} for each entanglement.}
    \label{T1}
\end{table}
\begin{figure}
\centering
\includegraphics[width=3.35 in]{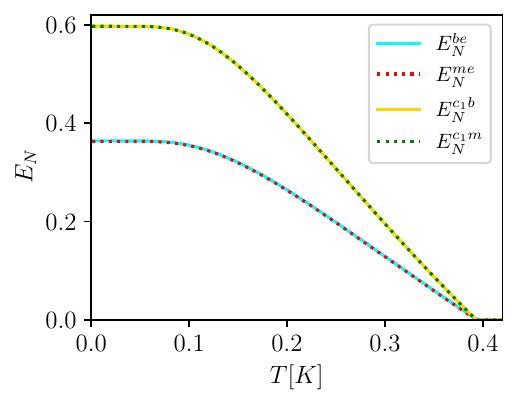}
\caption{The bipartite entanglements $E^{be}_N$, $E^{me}_N$, $E^{c_1b}_N$, and $E^{c_1m}_N$ against temperature $T[K]$. All the relevant parameters are mentioned in Table \ref{T1}.}
\label{f8}
\end{figure}
We have demonstrated that the CFL significantly enhances the degree of entanglement.
However, we have not discussed the robustness of entanglement against temperature in the presence of CFL.
In Fig. \ref{f8}, we plot distant bipartite entanglements, $E_{N}^{be}$, $E_{N}^{me}$, $E_{N}^{c_1m}$, and $E_{N}^{c_1b}$ as a function of temperature $T$ for the optimized set of parameters given in Table-\ref{T1}.
Previous study shows that the distant entanglement between phonon-ensemble and magnon-ensemble subsystems vanishes at $T\approx200$mK~\cite{PhysRevA.109.043708}.
Similarly, it was shown that the cavity-1 photon-phonon and cavity-1 photon-magnon entanglements survive up to $T\approx180$mK and $T\approx190$mK, respectively.
At these temperatures, our system exhibits much stronger entanglement than the maximum entanglement at zero temperature demonstrated in the previous study without CFL~\cite{PhysRevA.109.043708}.
Therefore, our results show that these entanglements are not only significantly increased at low temperatures, but their robustness against temperature is also increased by approximately twofold with $T\approx400$mK, due to the introduction of CFL. 
Finally, we present the effect of CFL parameters, $r_B$ and $\phi$, on the phonon-ensemble $E_N^{be}$ and magnon-ensemble $E_N^{me}$ entanglements in Fig.~\ref{f9}.
For $r_B=0$ and the given set of parameters, we have $E_N^{be}=E_N^{me}=\approx 0.1$.
In the presence of CFL, the entanglement values strongly depend on the strength $r_B$ along with the phase $\phi$ of the beam-splitter.


Here, we briefly outline the potential experimental feasibility of the proposed scheme.
The field of cavity magnomechanics has received considerable attention in recent years, with typical experiments involving 3D microwave cavities containing one or more YIG spheres~\cite{10.1126/sciadv.abe6379, PhysRevLett.130.193603}.
Besides, several other hybrid structures are also explored, including planar cavities, lumped-element cavities, feedback-coupled active cavities, microstrip, and superconducting architecture~\cite {ZARERAMESHTI20221, Yao2017, PhysRevLett.128.047701, PhysRevLett.132.206902}.
For instance, Li \textit{et. al.} recently demonstrated microwave-mediated remote strong magnon-magnon coupling using superconducting circuits~\cite{PhysRevLett.128.047701}.
Similarly, the coherent interaction between microwave fields with atomic ensembles and spin systems has been realized using superconducting coplanar waveguides~\cite{PhysRevLett.103.043603}.
Moreover, a microwave analog of a beam splitter is commonly used in such platforms~\cite{Pogorzalek2019}.
We believe a hybrid setup combining these established technologies can be used to realize our proposed scheme.
In our numerical simulations, we used experimentally feasible parameters typically realized in 3D microwave cavities.
Therefore, we believe that such a cavity coupled with superconducting transmission line resonators may be used to realize this scheme. 
The output of one port of such a cavity can be fed back to the input port using a microwave transmission line, forming a coherent feedback loop~\cite{PhysRevX.3.021013, PhysRevApplied.15.024056}.

\section{Conclusion}
In conclusion, we studied distant bipartite macroscopic entanglement of an atomic ensemble in cavity-1 with magnon and phonon modes of the YIG sphere in cavity-2.
We have introduced a CFL that injects the output of cavity-1 into cavity-1 via a beam splitter, thereby modifying its effective decay rate and detuning depending upon the reflectivity and phase shift introduced by the beam splitter.
We show that a particular choice of these parameters results in significant enhancement of all types of bipartite entanglement, including distant entanglement.
The CFL also significantly increases the parameter space where strong entanglement exists in comparison to the previous study of a similar system without CFL.
Similarly, the robustness of entanglement against temperature has increased up to $400$mK, which is two times higher than the previously reported temperature limit of a similar system.
Since atomic ensembles are often used as quantum memory nodes in quantum networks, we believe our proposal may have potential applications in quantum communications.

\section*{Acknowledgement}
We acknowledge the fruitful discussions with Bakht Hussain.

\bibliographystyle{apsrev4-2}

\bibliography{Ref.bib}

\end{document}